\documentclass[aps,prl,twocolumn,showpacs,floatfix,superscriptaddress,unsortedaddress]{revtex4}

\usepackage{amsmath,amssymb}
\usepackage{graphicx,color}
\usepackage{times}
\usepackage{soul}

\newcommand{\varH}{{\mathcal{H}}}
\newcommand{\bfS}{\mathbf{S}}%
\newcommand{\bfk}{{\mathbf{k}}}
\newcommand{\bfsigma}{{\boldsymbol{\sigma}}}
\newcommand{\vc}[1]{\mathbf{#1}}
\newcommand{\up}{{\uparrow}}
\newcommand{\down}{{\downarrow}}

\begin{document}

\title{The Two-impurity Anderson Model Revisited: Competition between Kondo
  Effect and Reservoir-mediated Superexchange in Double Quantum Dots}

\author{Minchul Lee}
\affiliation{Department of Physics, Korea University, Seoul 136-713, Korea}
\affiliation{Department of Applied Physics, Kyung Hee University, Yongin
  449-701, Korea}
\author{Mahn-Soo Choi}
\email{choims@korea.ac.kr}
\affiliation{Department of Physics, Korea University, Seoul 136-713, Korea}
\author{Rosa L\'opez}
\affiliation{Departament de F\'{i}sica, Universitat de les Illes Balears,
  E-07122 Palma de Mallorca, Spain}
\author{Ram\'on Aguado}
\affiliation{Teor\'{\i}a de la Materia Condensada,
  Instituto de Ciencia de Materiales de Madrid (CSIC) Cantoblanco,
  28049 Madrid, Spain}
\author{Jan Martinek}
\affiliation{Institute of Molecular Physics,
  Polish Academy of Sciences, Smoluchowskiego 17,
  60-179 Pozna\'n, Poland}
\author{Rok \v{Z}itko}
\affiliation{J. Stefan Institute, Jamova 39, SI-1000 Ljubljana, Slovenia}

\date{\today}

\begin{abstract}
  We study a series-coupled double quantum dot in the Kondo regime modeled by
  the two-impurity Anderson model and find a new conduction-band mediated
  superexchange interaction that competes with Kondo physics in the strong
  Coulomb interaction limit. Our numerical renormalization group results,
  complemented with the higher-order Rayleigh-Schr\"odinger perturbation
  theory, show that the novel exchange mechanism leads to clear experimental
  consequences that can be checked in transport measurements through double
  quantum dots.

% We study series-coupled double quantum dots in the Kondo regime
% using numerical renormalization group methods, focusing on the
% origin and influence of the inter-dot exchange interaction. In
% contrast to previous results, the conductance as a function of the
% inter-dot coupling surprisingly exhibits similar features for
% intermediate and large Coulomb repulsion, which are in both cases
% related to a quantum phase transition (cross-over) driven by the
% exchange interaction. Even for an infinitely strong repulsion, the
% inter-dot spin-spin correlations are strongly anti-ferromagnetic due
% to a new type of higher-order conduction-band-mediated exchange
% interaction.
% We study series-coupled double quantum
% dots in the Kondo regime using numerical renormalization group methods,
% focusing on the origin of the inter-dot exchange interaction and its effect on
% the conductance.
% %%
% We consider both intermediate and large Coulomb repulsion and show that the
% dependence of the % linear
% conductance on the inter-dot coupling unexpectedly,
% in contrast to previous result, in both cases exhibits common features related
% to the vicinity of a quantum phase transition (cross-over) driven by the
% exchange interaction. It can be explained by the observation that even in the
% limit of infinitely strong repulsion, the inter-dot spin-spin correlations are
% strongly anti-ferromagnetic due to a new type of higher-order
% conduction-band-mediated exchange interaction.
\end{abstract}

\pacs{73.21.La; 73.23.-b; 73.22.-f}
\maketitle

\begin{figure}[!t]
  \centering
  \includegraphics[width=5.5cm]{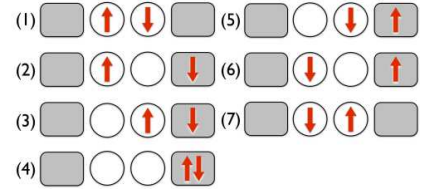}
  \caption{The sequence of six virtual processes of the most significant
    contribution to the conduction-band-mediated exchange $J_I$ in the
    large-$U$ limit. Here the circles (rectangles) denote dots (leads).}
  \label{fig:1}
\end{figure}

\begin{figure}[!t]
  \centering
  \includegraphics[width=6.5cm]{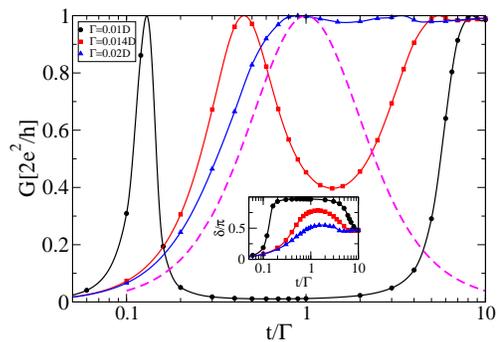}
  \caption{Zero-temperature linear conductance and phase-shift difference
    $\delta = \delta_e - \delta_o$ (inset) in the large-$U$ case ($U=\infty$).
    Lines are guidance for eyes. The dashed line is obtained from the SBMFT
    calculation~\cite{dqdsbmft}.}
  \label{fig:4}
\end{figure}

%\paragraph{Introduction.---}
Quantum dots (QDs)~\cite{qds} behave as quantum impurities, thus they exhibit
the Kondo effect, the most spectacular manifestation of which is the transition
from near-zero conductance due to Coulomb blockade to perfect
transmission~\cite{unitarylimit,qdexps} as the temperature is lowered well
below the Kondo temperature, $T_K$. More complex setups allow tailored
realizations of strongly correlated electron systems in new situations. Double
quantum dots (DQDs) \cite{dqdsbmft,expdqd2kIp1,expdqd2kIp2,dqdrkky}, for
instance, are a minimal system for studying a lattice of magnetic impurities in
a tunable environment. In the two-impurity Kondo model, the competition between
the Kondo effect and the anti-ferromagnetic (AF) interaction between impurities
leads to a second-order quantum phase transition (QPT)~\cite{2kIpa}. It is also
known that the QPT is replaced by a crossover if electrons can tunnel between
the impurities~\cite{2kIpb}. In series-coupled DQDs, the most important
contribution to the total AF exchange coupling is the superexchange coupling
$J_U\approx{4t^2}/{U}$, where $t$ is the inter-dot hopping and $U$ the on-site
interaction.
This superexchange vanishes ($J_U\to0$) as $U\to\infty$, which leads to the
common belief~\cite{dqdsbmft,dqdrkky} that \emph{the Kondo physics always
  prevail} in the large-$U$ limit. In this Letter, we present numerical
renormalization group (NRG) calculations which confute the above claim and
unambiguously demonstrate the presence of a new type of spin-spin interaction
which competes with Kondo physics in the large-$U$ limit. Our main findings are
illustrated in Fig.~\ref{fig:4} where the peak position of the linear
conductance through a DQD versus $t$ exhibits a striking and unexpected
dependence on $\Gamma$, the QD level broadening due to the coupling to the
reservoir. Interestingly, the novel spin-spin interaction reported here leads
to observable experimental consequences like zero conductance $G\approx 0$ in
regions where unitary transport $G\approx 2e^2/h$ is expected and vice
versa. Intuitively, we can understand this spin-spin interaction in terms of
virtual tunneling events involving conduction-band electrons; see
Fig.~\ref{fig:1}.
The strength of this spin-spin interaction, which we call
\emph{conduction-band-mediated superexchange}, $J_I$, can be estimated in the
large-$U$ limit using Rayleigh-Schr\"odinger perturbation theory (RSPT) as
\begin{equation}
  \label{eq:JI}
  J_I
  =
  \frac{4t^2\Gamma^2}{\pi^2}\int_{E_F}^{D}
  \frac{d\epsilon_ {1}d\epsilon_{2}}%
  {(\epsilon_{1}-\epsilon)^2
    (\epsilon_{2}-\epsilon)^2
    (\epsilon_{1}+\epsilon_{2}-2 \epsilon)}
\end{equation}
for $t\ll|\epsilon|$ where $\epsilon<0$ is the single-particle energy level of
QD. Here $E_F{=}0$ and $D$ are the Fermi energy and the half-width of the
conduction band, respectively. For a wide band ($|\epsilon|\ll D$), it
simplifies to $J_I \approx ct^2\Gamma^2/|\epsilon|^3$ with a constant
$c=8(1-\ln2)/3\pi^2\approx0.083$.  Interestingly, a single lead is sufficient
to induce a finite exchange $J_I\neq 0$, as seen in
Fig.~\ref{fig:1}. Remarkably, this high-order tunneling process is able to
affect the transport properties of DQDs. In what follows, we present in detail
our NRG study and discuss thoroughly the conditions under which $J_I$ can
compete with Kondo physics in regions where $J_U$ vanishes.

\paragraph{Model and methods.---}
We model the DQD by the two-impurity Anderson model:
\begin{multline}
  \label{eq:H}
  \varH
  =
  \sum_{\ell\bfk\mu}\epsilon_{\bfk} c_{\ell\bfk\mu}^\dag c_{\ell\bfk\mu}
  +
  \sum_\ell \left(\epsilon n_\ell + U n_{\ell\up} n_{\ell\down}\right)
  \\
  \mbox{}
  -t \sum_\mu \left(d_{1\mu}^\dag d_{2\mu} + \mathrm{h.c.}\right)
  +
  \sum_{\ell\bfk\mu} V \left[c_{\ell\bfk\mu}^\dag d_{\ell\mu}
  +
  \mathrm{h.c.}\right]
\end{multline}
Here $c_{\ell\bfk\mu}$ destroys a spin-$\mu$ electron with energy
$\epsilon_{\bfk}$ in the lead $\ell=1,2$, and $d_{\ell\mu}$ an electron in the
dot $\ell=1,2$; $n_\ell \equiv \sum_\mu d_{\ell\mu}^\dag d_{\ell\mu}$ is the
occupation of the dot $\ell$. For simplicity, we assume reflection symmetry
with respect to the interchange $1\leftrightarrow 2$ of leads and dots. The
single-particle energy $\epsilon$ on each dot is tunable by gate voltages.
% $U$ is the on-site Coulomb interaction.
The hybridization between the dot and lead is characterized by $\Gamma = \pi
\rho V^2$, with a flat-band density of states $\rho=1/2D$.  Throughout this
work, we fix $\epsilon = -0.1D$ and focus on the Kondo regime with localized
level, $\Gamma \ll |\epsilon|$ and large charging energy $U \ge 2|\epsilon|$.

We solve the Hamiltonian with the standard NRG
procedure~\cite{Wilson,Sakai92,Izumida97}.
% employing parity, spin, and charge quantum numbers to represent the
% eigenstates of the Hamiltonian~\cite{sakai,izumida}.
At low temperatures, a Fermi-liquid system is described by an effective
Hamiltonian that takes a form similar to the original Hamiltonian but with
renormalized parameters~\cite{renormalization}. Previous works have applied
this theoretical approach to deal with single dots~\cite{renormalization}. Here
we extend it to treat the two-impurity problem of a DQD.  It is then
technically convenient to change to the \emph{parity basis} (even and odd):
$c_{e(o)\bfk\mu}=(c_{1\bfk\mu}\pm c_{2\bfk\mu})/\sqrt{2}$ and
$d_{e(o)\mu}=(d_{1\mu}\pm d_{2\mu})/\sqrt{2}$.
In this basis, the Hamiltonian in Eq.~(\ref{eq:H}) reads as
\begin{multline}
  \varH
  =
  \sum_{s=e,o}
  \left[
    \sum_{\bfk\mu}\epsilon_\bfk c_{s\bfk\mu}^\dag c_{s\bfk\mu}
    +
    \epsilon_{s} n_s + \frac{U}{2} n_{s\up} n_{s\down}
  \right]
  \\
  \mbox{}
  + \frac{U}{4} n_e n_o - U \bfS_e\cdot\bfS_o
  - \frac{U}{2}
  \left[d_{e\up}^\dag d_{e\down}^\dag d_{o\down} d_{o\up} + h.c.\right]
  \\
  \mbox{}
  + \sum_{s\bfk\mu} V\left[c_{s\bfk\mu}^\dag d_{s\mu} + h.c.\right]
\end{multline}
%
% \begin{multline}
% \varH=\sum_{s=e,o} \{\sum_{\bfk\mu} \epsilon_\bfk c_{s\bfk\mu}^\dag
% c_{s\bfk\mu}+ (\epsilon_{s} n_s + \frac{U}{2} n_{s\up} n_{s\down}) \\{}
% + \sum_{\bfk\mu} V[c_{s\bfk\mu}^\dag d_{s\mu} + h.c.\}+ \frac{U}{4} n_e n_o- U
% \bfS_e\cdot\bfS_o \\{}
% - \frac{U}{2}[d_{e\up}^\dag d_{e\down}^\dag d_{o\down} d_{o\up} + h.c.]
% \end{multline}
% Here $n_s \equiv \sum_\mu d_{s\mu}^\dag d_{s\mu}$ and the spin operator is
% $\bfS_s = \frac12 \sum_{\mu\mu'} \bfsigma_{\mu\mu'} d_{s\mu}^\dag d_{s\mu'}$
% with Pauli matrices $\bfsigma$.
Here the local spin operator is defined by
\begin{math}
\bfS_s = \frac12 \sum_{\mu\mu'} \bfsigma_{\mu\mu'} d_{s\mu}^\dag d_{s\mu'}
\end{math}
with $\bfsigma$ being Pauli matrices, and the even (odd) level is given by
$\epsilon_{e(o)}=\epsilon \mp t$.

%%
% A Schrieffer-Wolff transformation of $\varH$ in the parity basis leads to the
% two-impurity Kondo Hamiltonian
% \begin{math}
% \varH_{K} =
% \sum_{s\in\{e,o\}}I_s\bfS_s\cdot\bfs_s(0) +
% J_{eo}\bfS_e\cdot\bfS_o
% \end{math}
% apart from the {electrode} part.  $\varH_{K}$ describes two localized
% spins, $\bfS_e$ and $\bfS_o$, that interact with each other via $J_{eo}$ as
% well as with the conduction spins $\bfs_e(0)$ and $\bfs_o(0)$ via $I_e$ and
% $I_o$, respectively.
%

%A Schrieffer-Wolff transformation of $\varH$
% leads to the
% two-impurity Kondo Hamiltonian
% \begin{math}
% \varH_{K} =
% \sum_{s\in\{e,o\}}I_s\bfS_s\cdot\bfs_s(0) +
% J_{eo}\bfS_e\cdot\bfS_o
% \end{math}
% apart from the {electrode} part,
% where $J_{eo}$ denotes interaction between localized spins, and $I_s$ with the conduction spins $\bfs_s$.

% $\varH_{K}$ describes two localized
% spins, $\bfS_e$ and $\bfS_o$, that interact with each other via $J_{eo}$ as
% well as with the conduction spins $\bfs_e(0)$ and $\bfs_o(0)$ via $I_e$ and
% $I_o$, respectively.

The renormalized parameters at the Fermi-liquid fixed point (which we denote by
appending an asterisk to bare parameter symbols) are extracted from the NRG
flow diagrams.  Additional terms, missing in the bare Hamiltonian, which are
allowed by the symmetry are generated in the renormalization process. In the
absence of the particle-hole (p-h) symmetry, for example, coefficients $U_s$ of
the term $n_{s\up}n_{s\down}$, hybridization $\Gamma_s$, and energies
$\epsilon_s$ will renormalize differently in each parity channel.
%%
%% Back in the local basis, $U^*_e-U^*_o$ corresponds to correlated hopping
%% between the dots 1 and 2, $\Gamma^*_e-\Gamma^*_o$ corresponds to hopping from
%% lead 1 to dot 2 and from lead 2 to dot 1, while renormalized energies
%% $\epsilon_s^*$ can be re-parametrized in terms of effective on-site energy and
%% hopping through $\epsilon_{e(o)}^* = \epsilon^* \mp t^*$.
In the p-h symmetric case we were able to determine all interaction terms in
the local basis, in particular, the exchange coupling $J^*_{12}$ between dot
spins $\bfS_1$ and $\bfS_2$. Finally, the linear conductance is determined by
even and odd quasi-particle scattering phase shifts as $G =
({2e^2}/{h})\sin^2(\delta_e-\delta_o)$~\cite{dqdsbmft}.  In the following we
consider two cases: (i) the model with the p-h symmetry and intermediate-$U$,
where the physics is dominated by $J_U$ and (ii) the large-$U$ case where $J_U$
is suppressed. Analysis of two cases reveals the emergence of a new type of
anti-ferromagnetic coupling $J_I$ generated by high-order virtual tunneling
events (see Fig.~\ref{fig:1}) among the electrons localized on dots and the
itinerant carriers in the reservoirs.
\begin{figure}[!t]
  \centering
  \includegraphics[width=7cm,clip=]{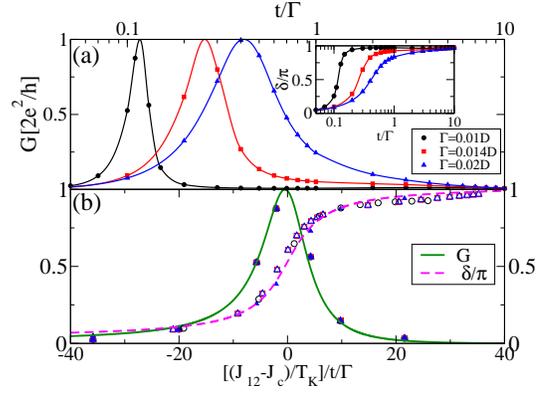}
  \caption{(a) Zero-temperature linear conductance and phase shift difference
    $\delta=\delta_e - \delta_o$ in the symmetric case (inset) as functions of
    $t/\Gamma$, $U=2|\epsilon|$. Lines are guidance for eyes. (b) Linear
    conductance $G$ (filled symbols) and phase shift as functions of the scaled
    exchange coupling, $(J_{12}-J_c)\Gamma/T_Kt$. Results for the phase shift
    differences and for conductance collapse well onto universal curves, the
    scaling functions fitted to the data.}
  \label{fig:2}
\end{figure}

\begin{figure}[!t]
  \centering
  \includegraphics[width=8cm,clip]{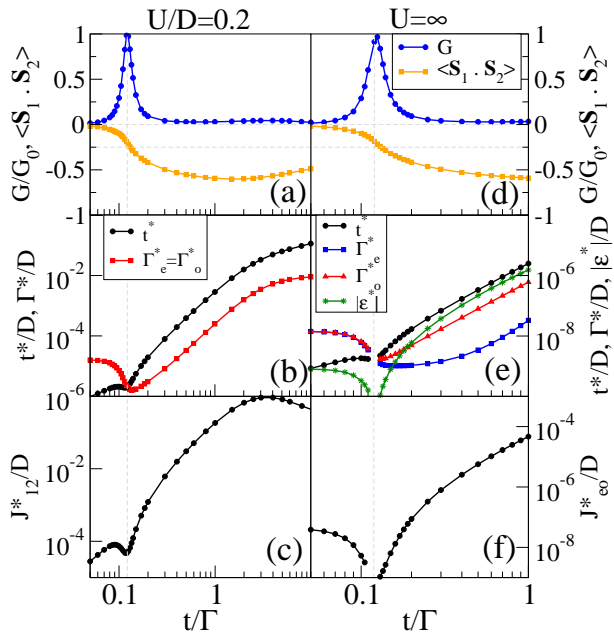}
  \caption{Zero-temperature properties of the p-h symmetric model (left panels)
    and the p-h asymmetric $U=\infty$ model (right panels) for
    $\Gamma=0.01D$. (a,d) Linear conductance and spin-spin correlation
    $\langle\vc{S}_1\cdot\vc{S}_2\rangle$. (b,e) Renormalized parameters,
    $t^*$, $\Gamma_e^*$, $\Gamma_o^*$, $|\epsilon^*|$. In the symmetric model,
    $\Gamma_e^*=\Gamma_o^*$ and $\epsilon^*=0$.  {(c,f) Renormalized
      superexchange coupling in the \emph{local basis} (c) and in the
      \emph{parity basis} (f).}}
  \label{fig:3}
\end{figure}

\paragraph{Particle-hole symmetric, intermediate-U regime.---}
Let us review briefly the main features for the p-h symmetric case in
Fig.~\ref{fig:2}. The Friedel-Langreth sum rule imposes the relation
$\delta_e+\delta_o=\pi$ for all $t$ and then the average occupation per spin
channel is one. The physics is governed by the competition between the Kondo
and the AF energy scales. For small $t$, $J_{12}\ll T_K$ and each dot spin
forms a Kondo singlet state with conduction electrons in the neighboring lead,
while the weak inter-dot coupling yields small conductance. Furthermore, the
spin-spin correlation $\langle \vc{S}_1\cdot \vc{S}_2\rangle$ vanishes; see
Fig.~\ref{fig:3}(a). As $t\to\infty$, $J_{12}\gg T_K$, and the dot spins are
locked into a local singlet state, thus inhibiting the Kondo effect, and
consequently the conductance is small in this limit as well. As $t$ increases
from 0 to $\infty$, the conductance exhibits one pronounced peak as shown in
Fig.~\ref{fig:2}(a) where the peak position is determined by the condition
$J_c(t=t_c) = J_{12}\sim T_K$. At this point the spin correlation takes the
value $\langle \vc{S}_1\cdot \vc{S}_2 \rangle=-1/4$ due to the singlet-triplet
degeneracy~\cite{2kIpa}. The crossover occurs from the Kondo to AF
regime~\cite{2kIpb,dqdsbmft} where $J_{12}\sim{}T_K$, and the parameters $t^*$
and $\Gamma^*$ become equal at $J_{12}=J_c$.  This crossover is known to be
well described by a scaling function~\cite{dqdsbmft}:
\begin{math}
  \delta/\pi = \phi\left[{(J_{12}-J_c)\Gamma/T_Kt}\right]
\end{math}
with $\phi(-\infty)=0$ and $\phi(\infty)=1$ which is in good agreement with our
numerical results; see Fig.~\ref{fig:2}(b).

\paragraph{Large-$U$ regime.---}
We found that the large-$U$ case is governed by spin-spin correlations mediated
by the conduction-band electrons.
% as demostrate unambiguously Figs.~\ref{fig:3}(d,e,f) and \ref{fig:4}.
The linear conductance features a plateau starting at $t \sim
|\epsilon|$~\cite{note1} and a peak at lower $t$; see
Fig.~\ref{fig:4}. Importantly, the peak reaches the unitary limit and shifts
toward larger $t/\Gamma$ with increasing $\Gamma$ as in the symmetric case.  It
must be emphasized that \emph{this behavior of the conduction peak is wholly
  unexpected and disagrees with previous analyzes for the $U\to\infty$
  limit}. We therefore continue with a more detailed study of this feature.  In
the large-$U$ limit, the usual superexchange $J_U \propto 1/U$ vanishes. For
this reason, the conductance peak has been traditionally ascribed to the
formation of even and odd Kondo states occurring at $\Gamma^*=t^*$. In
addition, some theories~\cite{dqdsbmft,nca} predicted that $\Gamma$ and $t$ are
renormalized in the same manner, thus the transition from individual Kondo
states to bonding and anti-bonding states was expected to happen at
$\Gamma=t$. However, the results in Fig.~\ref{fig:4} are in clear disagreement
with this picture: the conductance peak is not located at $\Gamma=t$, but
rather at $t<\Gamma$ and it is shifted toward smaller $t/\Gamma$ with
decreasing $\Gamma$. This suggests the presence of some processes which are
responsible for the different renormalization of $\Gamma$ and $t$.  We indeed
find an exchange coupling between the dot spins, which is manifested in the
unexpectedly strong anti-ferromagnetic correlation
$\langle\vc{S}_1\cdot\vc{S}_2\rangle<0$ in Fig.~\ref{fig:3}(d). At the
conductance peaks, this correlation takes the value of $-1/4$.  The origin of
the exchange coupling at $U \to \infty$ is illustrated in Fig.~\ref{fig:1}
in which the dot spins get to be locked into a singlet state by means of
virtual tunneling processes where conduction-band electrons participate.

%The phase shifts behave qualitatively as in the symmetric case up to $t \sim
%\Gamma$. For $t=0$, $\delta_e = \delta_o = \pi\langle n_e + n_o\rangle/4
%\lesssim \pi/2$, and as $t$ increases, $\delta_e$ grows toward $\pi$, while
%$\delta_o$ decreases; the sum of phase shifts is less than $\pi$ due to
%inhibited double occupancy {of single dots} which reduces the average
%occupation below one, Fig.~\ref{fig:3}(d). As $t$ increases beyond $\Gamma$,
%$\delta_e$ decreases to $\pi/2$, and $\delta_o$ increases slightly and then
%drops to zero. Consequently, the phase shift difference $\delta$, see
%Fig.~\ref{fig:4}(c), shows a non-monotonic behavior: starting from
%$\delta=0$ for $t=0$, it first increases rather rapidly and then decreases,
%eventually saturating around $\pi/2$. The linear conductance thus features a
%plateau starting at $t \sim |\epsilon|$ and a peak at lower $t$,
%Fig.~\ref{fig:4}(b). The peak reaches the unitary limit and it is shifted
%toward larger $t/\Gamma$ with increasing $\Gamma$ as in the symmetric case.
%It must be emphasized that this behavior of the conduction peak is wholly
%unexpected and it disagrees with previous results for the $U\to\infty$
%limit. We therefore continue with a more detailed study of this feature.

The strength of this conduction-band-mediated superexchange coupling $J_I$ in
the large-$U$ limit is estimated using the RSPT; see Eq.~(\ref{eq:JI}). The
RSPT (or projection method) should be enough to extract the almost true value
of $J_I$ since $J_I$ is due to the charge fluctuations, as $J_U$ is; see
Fig.~\ref{fig:1}. Further scaling would affect only the spin fluctuation
part related to the Kondo effect and not to the local spin interactions. Hence,
$J_I$ computed from the RSPT should be, at least in order of magnitude, very
close to the one that competes with the Kondo scale. The superexchange coupling
$J_I$ is generated at the fourth order of $V$ in the RSPT scheme and in the
limit $t\ll |\epsilon|, U$ it is of the second order in $t$; see
Eq.~(\ref{eq:JI}). It is remarkable that such higher-order processes lead to
\emph{sufficiently strong spin correlations to qualitatively affect the
  transport properties of the DQD}. The maximal conductance at the crossover
region occurs because $J_I$ drives the system from the Kondo phase to the
anti-ferromagnetic phase at $J_I \approx T_K$, in full analogy with the
symmetric case. This is underpinned by the observation that the renormalized
$\epsilon^*$ goes through zero, even and odd hybridizations become equal,
$\Gamma^*_e=\Gamma^*_o=\Gamma^*$, and $\Gamma^*=t^*$; see
Fig.~\ref{fig:3}(e). At this particular point, the p-h symmetry is,
surprisingly, restored at low energies. We also find that the magnetic
interaction $J^*_{eo}$ (it is technically difficult to extract $J_{12}^*$
unlike in the p-h symmetric case) again exhibits a local minimum at the cross
over [see Fig.~\ref{fig:3}(f)]. The fact that the conductance peak is shifted
away from $\Gamma=t$ can be regarded as the experimental proof of the existence
of a new type of spin-spin correlations not accounted so far.
Other theoretical techniques like slave-boson decomposition~\cite{dqdsbmft,nca}
do not describe properly these magnetic correlations between the dot spins that
arise naturally as the interplay between the tunneling events and the Coulomb
repulsion. In these approaches, spin exchange interactions are usually included
\emph{ad hoc}~\cite{Izumida97}. Our results suggest that both $J_U$ and $J_I$
have to be included in slave-boson schemes in order to achieve a good agreement
with NRG calculations.

\begin{figure}[!t]
  \centering
  \includegraphics[width=7cm]{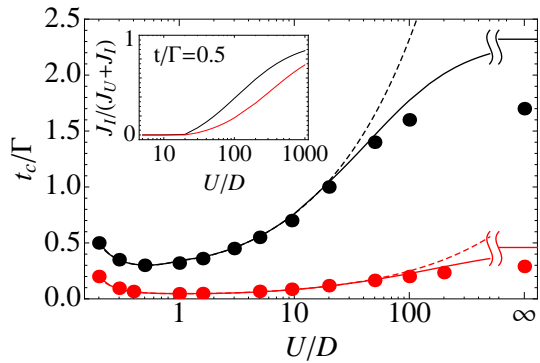}
  \caption{The conductance-peak position $t_c/\Gamma$ as a function of the
    Coulomb energy $U$ for different values of $\Gamma$: $\Gamma/D=0.012$ (red)
    and 0.02 (black). The dotted and solid lines are the peak positions
    estimated from the condition $J_U=2.2T_K$ and $J_U + J_I=2.2T_K$,
    respectively. Inset: the relative strength of $J_I$ with respect to the
    total superexchange coupling.}
  \label{fig:5}
\end{figure}

\paragraph{Crossover.---}
Now we address the intermediate regime with medium $U$ where both
anti-ferromagnetic couplings are finite.  After a lengthy calculation it is
possible to obtain an analytical expression for $J_I$ as a function of
$U$~\cite{note2}. The expression is too cumbersome to appear here and we just
plot the final results.  The inset in Fig.~\ref{fig:5} shows the relative
portion of $J_I$ in the total superexchange versus $U$. For sufficiently large
$U$, $J_I$ surely dominates over $J_U$ that eventually vanishes as
$U\to\infty$.  For intermediate values of $U\lesssim 10D$, $J_I\ll J_U$ the
physics is governed by the superexchange interaction $J_U$, while it is $J_I$
that suppresses the Kondo correlations for larger $U$. Since $J_I$ is
proportional to $\Gamma^2$, the value of $U$ where $J_I$ becomes dominant over
$J_U$ can be lower with increasing $\Gamma$.  The competition between $J_I$ and
$J_U$ are also reflected in the main plot of Fig.~\ref{fig:5}, where the linear
conductance peak position versus $U$ is plotted. While for intermediate $U$ the
peak position is well identified by the condition $J_U\sim T_K$ (dotted lines),
which is in agreement with previous theories, the deviation becomes significant
for large $U$, and in the $U\to\infty$ limit the peak position is given by the
condition $J_I\sim T_K$. Interestingly, a good fitting for arbitrary value of
$U$ is obtained by determining the peak position by imposing the condition
$J=J_I+J_U=2.2 T_K$, with $T_K = \sqrt{\Gamma \min(U,D)/2} \exp[2\pi
\epsilon(1+\epsilon/U)/\Gamma]$. Thus, the conduction-band-mediated
superexchange explains in a very suitable manner (both qualitatively and
quantitatively) the saturation of the conductance peak as $U$ is increased. Our
estimation of the peak position from the RSPT is slightly larger than one
obtained from the NRG calculations. It is due to our neglect of further scaling
of $J_I$ with integration of the conduction band and the renormalization of
$T_K$ due to the inter-dot coupling. Both renormalizations should be quite
marginal and may lead to slightly larger $J_I$ and smaller $T_K$, with which
the condition $J=2.2T_K$ gives rise to smaller conductance peak position.
% Probably, the condition to extract the conductance peak position
% could be improved by using a different Kondo temperature.  For large-$U$ the
% spin singlet and triplet states become almost degenerate, which may give rise
% the S=1 Kondo effect with different Kondo temperature.

\paragraph{Discussion.---}
The RSPT estimation predicts that $J_I$ becomes comparable to $J_U$ at
$U\approx U_c = 4|\epsilon|^3/c\Gamma^2$. The ratio $U_c/|\epsilon|$ is then of
the order of a few hundreds in order to retain the Kondo effect
$(|\epsilon|/\Gamma\ll1)$. The condition can be satisifed through the electric
control of the single-particle level in tunable samples, or by making use of
ultrasmall samples with strong Coulomb interaction. One of the latter
canditates is a junction made of molecular dimers where $U$ is about a few
eV. While the interdot coupling of dimers can hardly be tuned, the gate voltage
control of $\epsilon$ and $\Gamma$ can nevertheless fulfill the crossover
condition $J_I\sim T_K$, allowing to examine the role of the superexchange on
the transport.

\paragraph{Conclusion.---}
We have reported on a new conduction-band mediated superexchange interaction,
$J_I$, that competes with Kondo physics in the large-$U$ limit of the
two-impurity problem. The novel exchange mechanism brings more unity to this
problem in terms of magnetic correlations, and the transport through the DQD
can be analyzed in terms of the competition between the Kondo correlation and
the anti-ferromagnetic interaction for all values of $U$. Transport experiments
can confirm the presence of high-order superexchange coupling by examining the
dependence of the peak position on the inter-dot hopping amplitude.

\begin{acknowledgments}
  R.A.\ acknowledges financial support from grant MAT2006-03741 (MEC-Spain) and
  CCG08-CSIC/MAT-3775.  M.-S.C.\ was supported by the grant R11-2000-071, the
  BK21, and the KIAS; and J.M.\ by the Polish grant for science in years
  2006-2009 as a research project; and R.L by the grant FIS2008-00781
  (MEC-Spain).
\end{acknowledgments}


\begin{thebibliography}{99}
\bibitem{note1} The conductance plateau for large $t$ in Fig.~\ref{fig:4} is
  due to one-electron $S=1/2$ Kondo effect. The one-electron even-orbital state
  is lowered below the two-electron states when $t \gtrsim |\epsilon|$.  Hence
  the conductance plateau is attributed to the formation of a single-impurity
  Kondo state in the even channel, and the conductance therefore reaches the
  unitary limit at zero temperature, independent of the value of $t/\Gamma$.
\bibitem{note2}
  Explicit expression for $J_I(U)$ will be given elsewhere.
\end{thebibliography}
\end{document}